\documentclass[doublecol]{epl2} 

\title{Scaling laws and universality in the choice of election candidates}

\author{M. C. Mantovani\inst{1,2} \and H. V. Ribeiro\inst{1} 
          \and M. V. Moro\inst{1} \and S. Picoli Jr.\inst{1} \and R. S. Mendes}
\shortauthor{M. C. Mantovani \etal}

\institute{                    
  \inst{1} Departamento de F\'isica and National Institute of Science and Technology for
              Complex Systems, Universidade Estadual de Maring\'a, 
              Av. Colombo 5790, 87020-900, Maring\'a, PR, Brazil\\
  \inst{2} Universidade Tecnol\'ogica Federal do Paran\'a, Campus 
              Campo Mour\~ao, 87301-006, Campo Mour\~ao, Paran\'a, Brazil
}


\abstract{
Nowadays there is an increasing interest of physicists in finding regularities related to social phenomena.
This interest is clearly motivated by applications that a statistical mechanical description
of the human behavior may have in our society. By using this framework, we address this work to cover an open question
related to elections: the choice of elections candidates (candidature process). Our analysis
reveals that, apart from the social motivations, this system displays features of traditional
out-of-equilibrium physical phenomena such as scale-free statistics and universality.
Basically, we found a non-linear (power law) mean correspondence between the number of candidates 
and the size of the electorate (number of voters), and also that this choice has a multiplicative 
underlying process (lognormal behavior). The universality of our findings is supported by data from $16$ 
elections from $5$ countries. In addition, we show that aspects of network scale-free can be connected
to this universal behavior.
}

\begin{document}
\maketitle

\section{Introduction}
Social phenomena are nowadays ubiquitous in the research performed by many physicists.
In these investigations, the human behavior plays a central role and it constitutes the
basic ingredient of the emergent picture. However, despite the complex scenario
related to human activities, statistical physics models have been successfully applied 
to explain collective aspects of social systems~\cite{Gonzalez1,Ratkiewicz,Ausloos,Rozenfeld,
Cattuto,Rybski,Radicchi,Ribeiro}. This success gives rise to the possibility that,
similarly to large-scale physical thermodynamic systems, large groups of interacting humans
may exhibit universal statistical properties. For the society organization in general, this statistical 
mechanical description of the human activities seems to be promising in resource management, 
service allocation, political strategies, among others.


In this context, aspects related to the formation and spreading of opinions have attracted great
interest. However, while several models~\cite{Castellano} such as the voter~\cite{Clifford}, Axelrod~\cite{Axelrod}, majority~\cite{Galam}
or the Sznajd~\cite{Sznajd} have been proposed aiming to investigate this scenario, far less attention
has been paid to empirical investigations. An exception is the electoral process which
has been extensively studied concerning the election results. In fact, 
Costa Filho \textit{et al.}~\cite{CostaFilho} (see also Ref.~\cite{CostaFilho2}) reported
that the distribution of votes among candidates for one Brazilian federal election follows
a power law in a limited interval of the number of votes. Bernardes \textit{et al.}~\cite{Bernardes} reproduced this behavior
employing a network model coupled with the Sznajd rule. By using Brazilian results of federal
and local elections Lyra \textit{et al.}~\cite{Lyra} showed that the distribution of votes is
well adjusted by a generalized Zipf's law.
Gonzalez \textit{et al.}~\cite{Gonzalez} found a similar behavior for Indian elections.
Travieso and Costa~\cite{Travieso} investigated a network model which qualitatively reproduces
the distribution profile of votes. Araripe \textit{et al.}~\cite{Araripe} studied the vote 
percentage in mayoral elections taking the number of candidates into account. 
For the case of
proportional elections Fortunato and Castellano~\cite{Fortunato} reported that
a universality class of vote distribution emerges when considering the party 
and the personal votes. The distribution profile is quite similar to a lognormal
and  was modelated by a branching process.
Correlations and memory were reported by Andresen \textit{et al.}~\cite{Andresen} for
Norwegian polling time series and Hern\'andez-Salda\~na~\cite{Hernandez}
investigated the vote distributions for one mexican party. Araripe and Costa Filho~\cite{Araripe2}
showed that the universality class of Fortunato and Castellano is not fully verified for
the Brazilian elections. Borghesi and Bouchaud~\cite{Borghesi} reported 
a logarithmic decay of the spatial correlations for turnout rates and winning fraction votes
with the distance among towns. Finally, in a recent article, Araujo \textit{et al.}~\cite{Araujo} 
proposed a model to investigate plurality elections.

Naturally, two aspects are essentials in an election: $i)$ have a prior set of candidates (candidature process)
$ii)$ from which some of them are elected by voters. This last aspect was investigated in 
Refs.~\cite{CostaFilho,CostaFilho2,Bernardes,Lyra,Gonzalez,Travieso,Araripe,Fortunato,Andresen,Hernandez,Araripe2,Borghesi,Araujo}. 
However, the candidature process, which is intrinsically related with the final results of an election since it drastically limits
the choice of voting, has not been investigated yet. 
In this letter we fill this hiatus showing that, despite social or cultural or even psychological characteristics, a robust regularity
emerges. In the following we present our data set as well as
some particularities of each analyzed election, the statistical analysis of the data and results
from our model, and our conclusions.


\begin{figure}[!t]
\centering
\includegraphics[scale=0.45]{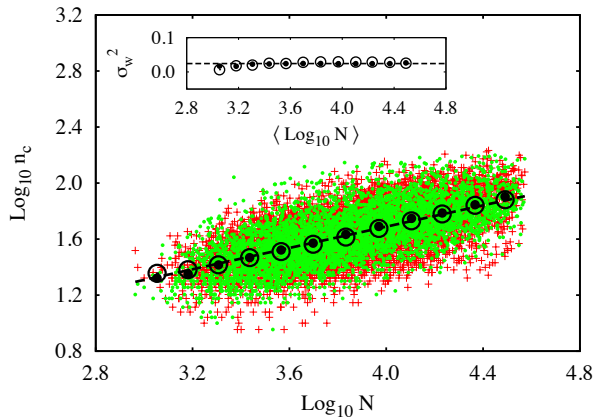}
\caption{(Color online) Scatter plot of the number of voters $N$ versus the number of candidates $n_c$
on base-10 logarithm scale for the 2008 Brazilian city council election. The red crosses represent
the empirical data and the green dots are simulation results obtained from the BA model with
$c=1.12$, $\sigma=0.196$ and $\beta=0.32$. The open (filled) circles are the averaged values for data 
(simulated results) binned logarithmically and the dashed line is a linear fit
to the data, where we found $n_c\sim N^{0.36}$. The inset shows the same comparison for the variance
values $\sigma_w^2$.
}\label{fig:cloud}
\end{figure}

\section{Data presentation and analysis}

As database, we have 16 elections~\cite{Sites} from Brazil (1996, 2000, 2004, 2008 - mayoral and city council elections),
Italy (2010 - mayoral elections), England (2006, 2010 - city council elections), Canada (2006, 2008 - parliamentary elections), and Australia (2004, 2007, 2010 - parliamentary elections).
There are 5 mayoral elections, 6 city council elections and 5 national elections. 
The data are constituted by the number of candidates in each local political division
as well as the number of voters. In Brazil, general elections are held every four years
and each city votes for executive (mayor) and legislative (councilor). The executive
is composed by one mayor and the number of legislative representatives is limited
by the size of the local populations. Therefore we have taken only towns with
up to 47600 inhabitants, for which the number of councilor is 9. They correspond
to $90\% \;(\sim5000)$ of the Brazilian cities and the non-use of this limitation does not affect
our results. For the Italian elections we have only mayoral elections (Provinciali and Comunali).
In the case of English elections, we have taken the council elections 
standing in all 624 wards in the London boroughs, where most of the wards have 3
seats (only 9 wards have a smaller number). In the case of Canada and Australia
the data are from the ``House of Commons'' and the ``House of Representatives'' elections,
each one having 308 and 160 members respectively. In both cases, the candidates
are elected in single-member elections disputed in each electoral district.

We start by investigating how the number of candidates ($n_c$) increases
with number of voters ($N$). In Fig.~\ref{fig:cloud} we show the scatter plot
of these two variables taking the base-10 logarithms for the 
2008 Brazilian city council election into account. Notice that a clear tendency emerges
suggesting that $n_c$ increases with $N$ following a power law, i.e., $n_c\sim N^\alpha$.
The linear behavior (in a log-log plot) is evidenced by the Pearson correlation 
coefficient $r=0.63$. This value of $r$ as well as the scatter plot also 
reveals that this power law correspondence is subjected to fluctuations. A possible
method to overcome the fluctuations is constructing windows logarithmically
spaced in $N$ and evaluating the average value of the inside points. These
mean values are also shown in Fig. \ref{fig:cloud} represented by open circles
and empirical results showed to be very robust with respect to the number of windows $w$. 
Now, the power law relation becomes evident where we have $\langle \log_{10} n_c \rangle 
= A + \alpha \,\langle \log_{10} N \rangle$ with the brackets representing the averages
and $\alpha=0.36$ for this election. 

\begin{figure*}[!t]
\centering
\includegraphics[scale=0.45]{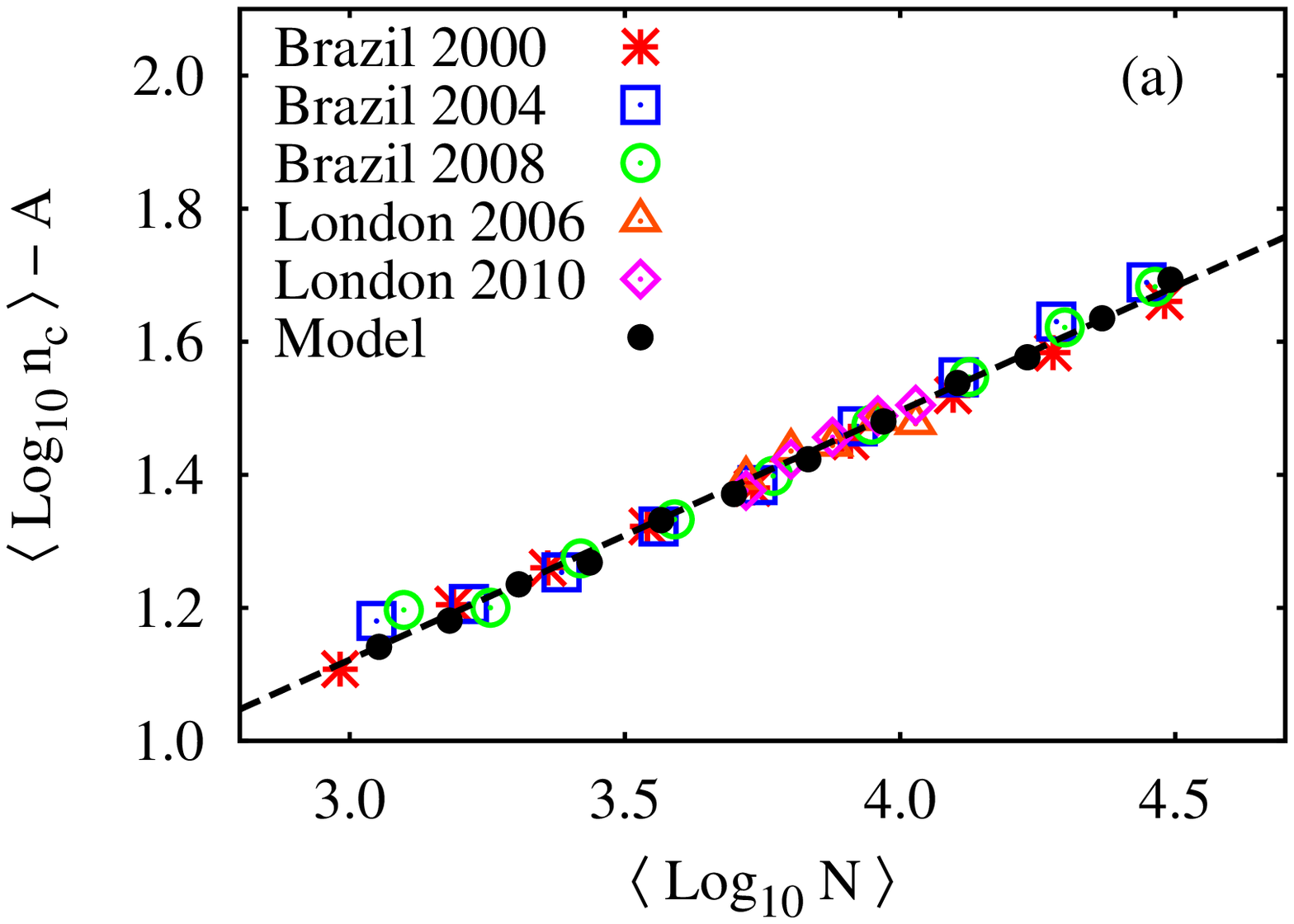}
\includegraphics[scale=0.45]{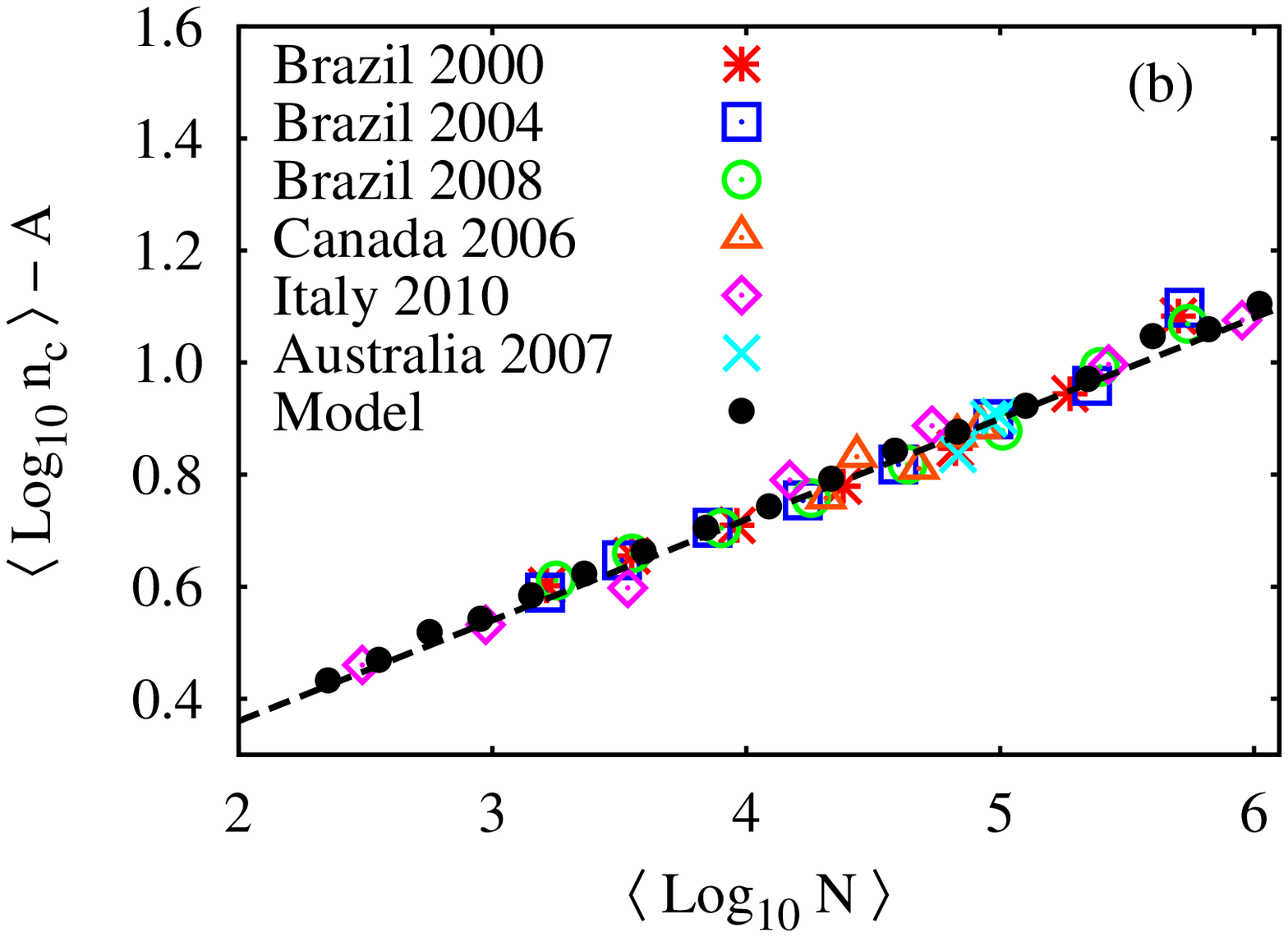}
\includegraphics[scale=0.45]{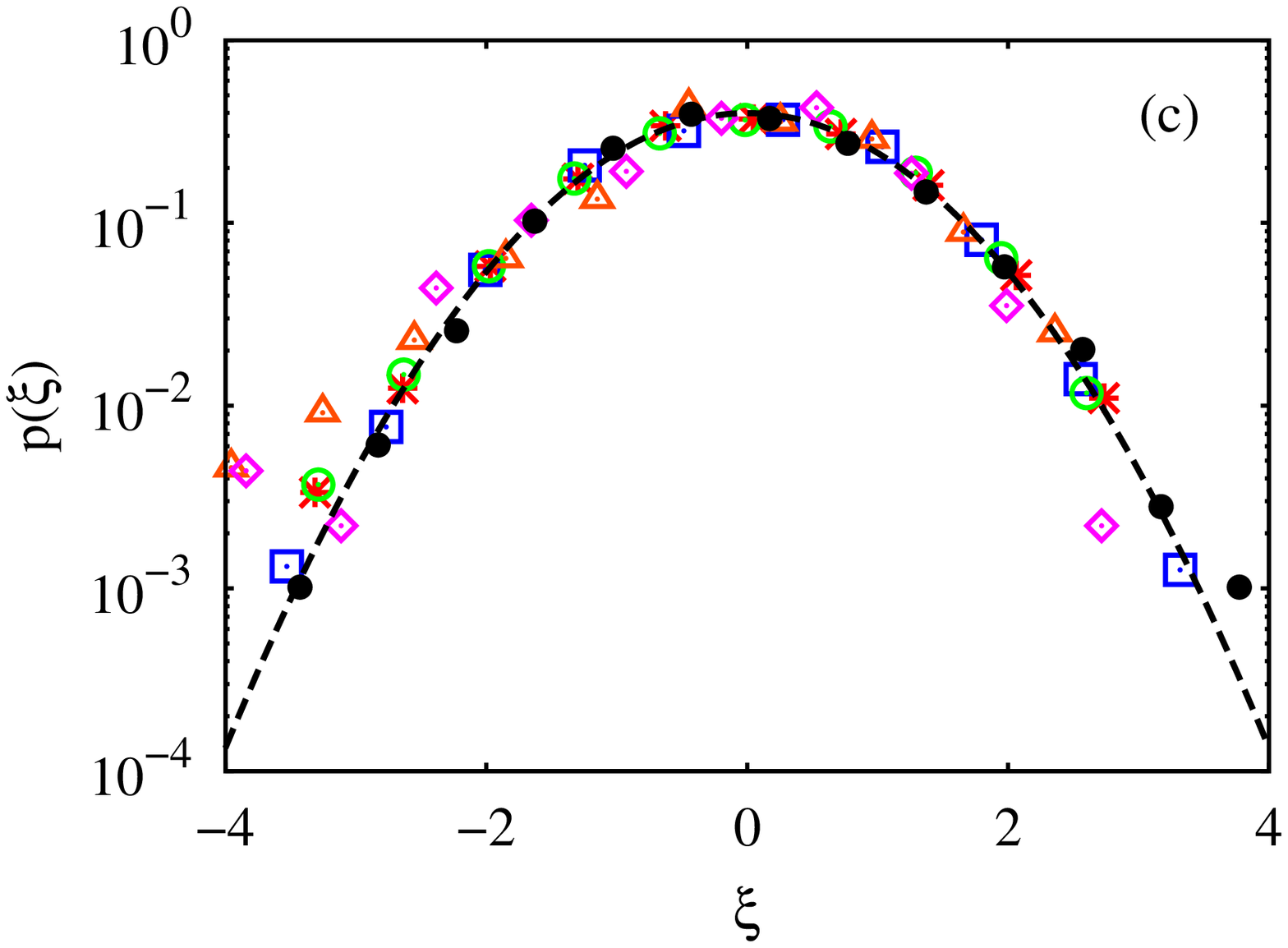}
\includegraphics[scale=0.45]{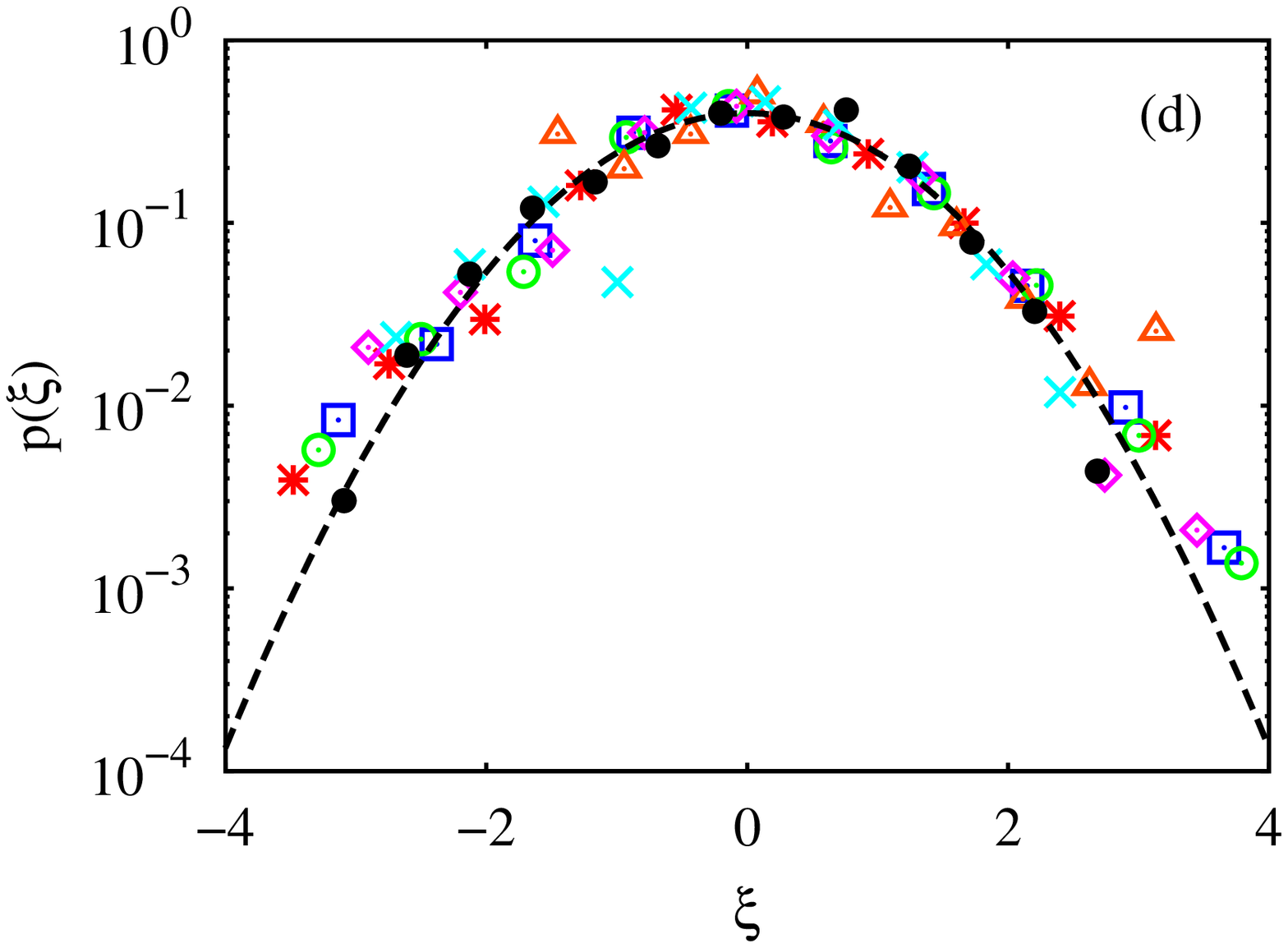}
\caption{(Color online) The upper panel shows the mean values of the logarithm of the number of candidates $n_c$ vs. number of
voters $N$ discounted the $A$ constant for (a) 5 multi-member elections and (b) 6 single-member elections. The 
dashed lines represent the mean value of the exponent $\alpha$, where we have found $\alpha=0.36$ for the
multi-member elections and $\alpha=0.18$ for single-member elections. In the bottom panel the 
fluctuation distributions for (c) the same multi-member elections of Fig.(a) and in (d) for the same single-member 
elections of Fig.(b) are showed. The dashed lines are Gaussian PDFs with zero mean and unitary variance. 
The different symbols represent the data and the black circles are the predictions of the network model, considering
the BA model with $\beta=0.32$ ($\beta=0.41$) for the multi-member (single-member) election. The other elections
present the same pattern.}\label{fig:comp}
\end{figure*}

Figure \ref{fig:comp}(a) shows these mean values for $5$ city council elections
and Fig. \ref{fig:comp}(b) does the same for $4$ mayoral and $2$ parliamentary elections.
Note that we have ploted these relations discounting the constant value $A$ aiming to collapse
the data. The good quality collapse and the values of $\alpha$ suggest two classes of universality: 
one for single-member elections characterized by $\alpha \approx 0.18$ and another for 
the city council elections (multi-member elections) with $\alpha \approx 0.36$. 
The previous result indicates that a voter also ponders the number of available seats in their
candidature process. However, it is interesting to note that for London boroughs election
there are 3 seats for each ward while in the Brazilian city council elections 9 seats are
disputed and nevertheless the value of the exponent $\alpha$ is approximately the same.
This suggests that the responsibilities of public position have also a central role in the candidature
process since the mayor or parliamentarian functions are in a different plateau of those ones
of a local representative. Moreover, local candidates may have only a ``short-range''  popularity 
restricted to their community while mayors or a parliamentarians should have
a greater influence.

Let us now address the fluctuation question by first considering the variance of the logarithm of the 
number of candidates $\sigma^2_w$. We evaluate the variance by employing the same procedure 
used for the mean values, i.e., constructing windows logarithmically spaced. Our findings basically indicate
that the variance does not depend on $N$, fact that can be observed from the inset of Fig. \ref{fig:cloud}.
In addition to the variance, we may also investigate the fluctuations around the power law mean relation. For this,
consider the variable $\xi=\frac{\log_{10} n_c - f_w(N)}{\sigma_w}$ where $f_w(N) = A+\alpha\,\langle\log_{10} N\rangle$ represents the
function adjusted to averaged values considering $w$ windows. Figures \ref{fig:comp}(c) and \ref{fig:comp}(d) show the probability
distribution function (PDF) of $\xi$ for the same elections of figures \ref{fig:comp}(a) and \ref{fig:comp}(b).
From these figures we observe that $\xi$ follows very close to the standard Gaussian. Moreover, we also found that,
like the mean values, the distributions does not depend on the number of windows $w$. 

At this point, it is interesting to summarize our empirical findings through the following expression:
\begin{equation}\label{eq:pe}
n_c= a \,\zeta(t)\, N^\alpha\,,
\end{equation}
or equivalently $\log_{10} n_c = \log_{10} a + \alpha \log_{10} N + \log_{10} \zeta(t)$  where 
$\log_{10} a = A$ and $\log_{10} \zeta(t)=\xi(t)$. Here, we use the variable $t$ to indicate
that $\zeta(t)$ and $\xi(t)$ are stochastic-like variables. While $\xi(t)$ is normally distributed, 
$\zeta(t)$ should be distributed according to a lognormal PDF. Therefore, Eq.(\ref{eq:pe})
represents a stochastic process where the number of candidates $n_c$ follows a power law mean
relation with $N$ driven by a multiplicative stochastic noise following a lognormal PDF. Naturally, 
simulations using Eq.(\ref{eq:pe}) are amazingly similar to the empirical ones. It is worth mentioning that, by
averaging this equation, we have $\langle \log_{10} n_c \rangle = A + \alpha \langle \log_{10} N \rangle$
since $\langle \log_{10} \zeta(t) \rangle = \langle \xi(t) \rangle =0$.

{A possible direction to go beyond Eq.(\ref{eq:pe}) is to employ the complex networks formalism~\cite{Albert}. 
Within this framework, the social agents are represented by nodes in a graph and the interactions or relationships are 
expressed by links among them. In our case, the social agents are the voters and the connections among them
represent a kind of desire to be a candidate. In a first approximation, we may directly relate the degree number
(number of links) to a kind of popularity measure, as more links (large degree) a voter has, the larger their popularity 
is and, consequently, larger is their probability to be a candidate (see Fig. \ref{fig:network}). 
{Naturally, other networks investigations and measurements could have been employed such as those ones
related to centrality~\cite{Borgatti} but, as minimal model, we focus only on the degree distribution $p(k)$.
Moreover, it is reasonable to consider that one voter will becomes a candidate if his popularity (degree) exceeds a characteristic value $k_c$. Consequently, the number of candidates existing in a given electorate $N$  
will be $n_c \sim N \int_{k_c}^{\infty} p(k) dk$. At this point it is interesting to remark that there is
no reason the suppose that $p(k)$ also sensibly depends on $N$ and thereby, the only way to $n_c$ behaves non-linearly
is to consider that the threshold value $k_c$ is a function dependent on $N$. 
In fact, regarding our data, we do not have any information related to the connections existing 
among the voters in such way that we can not indicate a specific network to represent the system.

On the other hand, a remarkable fingerprint of many social networks is the scale-free distribution 
of the degree number, in other words, the distribution of the number of links (degree) is often well 
described by a power law $p(k)\sim k^{-\gamma}$. This universal feature is usually related to the existence 
of preferential attachment inside social networks~\cite{Barabasi} and has been found in several
social systems (for examples see table II of Ref.~\cite{Albert}).
Thus, in this context, a scale-free network where $p(k)\sim k^{-\gamma}$ sounds as a natural 
choice for model our social system. Therefore,  $n_c \sim N k_c^{-\gamma+1}$ and since we empirically know that $n_c\sim N^\alpha$, the characteristic 
value $k_c$ should depend on the number of voters $N$ following another power law, i.e., $k_c \sim N^\beta$ with 
$\beta=\frac{\alpha-1}{1-\gamma}$. Of course that candidates are not always popular persons, since in general there is
no restrictions to be a candidate. In this way, $k_c$ may not be purely deterministic which can be accomplished
employing that $k_c = c \,\psi_\sigma(t)  N^\beta$ where $c$ is positive constant and
$\psi_\sigma(t)$ is a random number with unitary mean and standard-deviation $\sigma$ (negative numbers are avoided).
At this point, it is worth mentioning that other choices for $p(k)$ will lead to other  functional dependences for $k_c$,
which are generally more complicated. For instance, when considering the Erd\"os-R\'enyi model~\cite{Erdos} 
(or random network) for which $p(k)\simeq \frac{e^{-\langle k \rangle} \langle k \rangle^k}{\Gamma(k+1)}$,
$k_c$ should be obtained from the transcendental equation $a N^{\alpha-1}=\int_{k_c}^{\infty} \frac{e^{-\langle k \rangle} \langle k \rangle^k}{\Gamma(k+1)}dk$.
}

\begin{figure}[!t]
\centering
\includegraphics[scale=0.29, angle=90]{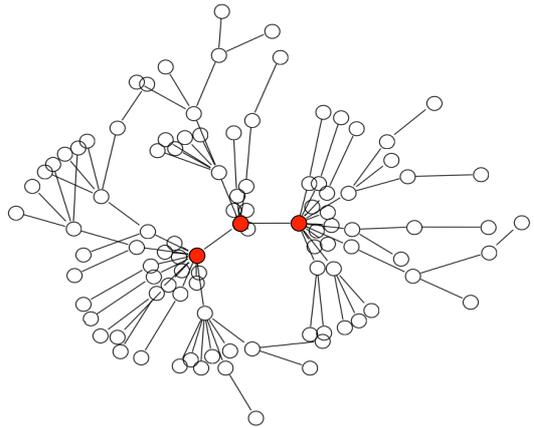}
\caption{(Color online) A schematic illustration of the network model: each vertex of the graph represents a voter
and the links among them are the social relationships, e.g., friendships. One voter ponders his popularity (here, the number of links)
in the community leading or not to the candidature. In this network, the highlighted nodes are the most popular ones
and consequently potential candidates.
}\label{fig:network}
\end{figure}

To be more specific, we investigate the above scenario employing the Barab\'asi and Albert~\cite{Barabasi} 
scale-free network (BA model) for which $\gamma=3$\cite{Barabasi,Barabasi2}. In our simulations, we constructed networks where the numbers of nodes are identical to the empirical values of $N$ and the $\beta$ values were obtained directly from the $\alpha$ values. 
The parameters $c$ and $\sigma$ should be adjusted for each election by minimizing, for example, the difference between the simulated 
and empirical values of the quantities $\langle n_c \rangle$ and $\langle \sigma_w^2 \rangle$, as we show in Fig. \ref{fig:cloud}.
We have found that $\sigma_w^2$ is always a small number (smaller than 1) and for the single-member elections this parameter
does not need to be considered, since the intrinsic fluctuations of the BA model are able to reproduce $\langle \sigma_w^2 \rangle$
and the PDFs, {indicating that the BA model can mimic the multiplicative noise existing in the data}. It is also interesting to note that both parameters are not important when employing the scaled variables of Fig. \ref{fig:comp}(a)-(d). From these figures we can observe that the model reproduces the scatter plot very well, as well as the mean values 
and PDFs.

\section{Summary and Conclusions}
To sum up, we investigated how the number of candidates ($n_c$) is related with the number of voters ($N$) of each electoral district. 
Analyzing data from 16 elections we found that $n_c$ increases with $N$ following a power law mean relation, $n_c\sim N^\alpha$.
The exponent $\alpha$ was found to be divided in two classes: $\alpha\approx 0.18$ for single-member elections and
$\alpha\approx 0.36$ for multi-member elections. We have also examined the fluctuations existing along the average relation
revealing that the candidature is driven by a multiplicative noise following a lognormal PDF. A scale-free network model 
representing the social
relationships was employed to reproduce the data behavior. The model results agree very well with the empirical data. 
The findings showed to be robust in the sense that they are independent of complex situation related to
social and cultural differences among the countries as well as of individual psychological attributes of the candidates. 
This robustness is frequently found in out-of-equilibrium systems or systems poised at criticality, 
where few mechanisms can lead to scale-free statistics and self-organized criticality\cite{Bak}.
Finally, we believe that our findings may have implications on the political scene concerning the 
candidature process. In particular our work shows that popularity-like measures may depend on the system size.

\acknowledgments
We thank CNPq and CAPES (Brazilian agencies) for financial support.

\end{document}